\documentclass[twocolumn,showpacs,prl]{revtex4}

\usepackage{graphicx,amsmath,amsfonts}  % required: for figures
\usepackage{multirow}
\graphicspath{{figs/}} % paths to look for figures

\usepackage{color}
\usepackage{float}
\begin{document}

\title{Demonstration of Turnstiles as a Chaotic Ionization Mechanism in Rydberg Atoms}

\author{Korana~Burke} 
\affiliation{ School of Natural Sciences, University of California, Merced, CA 95344, USA}
\author{Kevin~A.~Mitchell}
\affiliation{ School of Natural Sciences, University of California, Merced, CA 95344, USA}
\author{Brendan~Wyker}
\affiliation{Department of Physics, Rice University, 6100 Main Street, Houston, Texas 77005, USA}
\author{Shuzhen~Ye}
\affiliation{Department of Physics, Rice University, 6100 Main Street, Houston, Texas 77005, USA}
\author{F.~Barry~Dunning}
\affiliation{Department of Physics, Rice University, 6100 Main Street, Houston, Texas 77005, USA}

\begin{abstract}
  We present an experimental and theoretical study of the chaotic ionization of quasi-one-dimensional potassium Rydberg wavepackets via a phase-space turnstile mechanism.  Turnstiles form a general transport mechanism for numerous chaotic systems, and this study explicitly illuminates their relevance to atomic ionization.  We create time-dependent Rydberg wavepackets, subject them to alternating applied electric-field ``kicks'', and measure the electron survival probability. Ionization depends not only on the initial electron energy, but also on the phase-space position of the electron with respect to the turnstile --- that part of the electron packet inside the turnstile ionizes after the applied ionization sequence, while that part outside the turnstile does not.  The survival data thus encode information on the geometry and location of the turnstile, and are in good agreement with theoretical predictions.
\end{abstract}

\pacs{32.80.Rm, % Rydberg states--excitation and ionization of atoms
05.45.Gg, %Chaos--applications of
05.45.Ac, %Chaos--low-dimensional
45.50.Pk %Particle Orbits--classical mechanics
}

\maketitle

Chaotic behavior appears in diverse complex systems, over an enormous range of physical scales, including the formation of weather patterns, mixing of fluids, firing of neurons, and transport in the solar system.  Among these, photoabsorbtion and ionization in atomic gasses have proven to be excellent testbeds for both classical and quantum chaos.  For example, oscillations in the photoabsorption spectra of atoms in applied fields have been intimately linked to chaotic electron orbits~\cite{Kleppner01}, and resonant islands have proven to be barriers to microwave ionization~\cite{Koch95}.  More recently, such resonant islands have been used to trap, control, and engineer electronic Rydberg wave packets~\cite{Dunning06}.  Such experiments highlight the utility of highly excited Rydberg electrons as high resolution probes of chaotic phase spaces.

Whereas the above examples focused on steady state, or nearly steady state, dynamics, the present work focuses on revealing the time-dependent mechanism underlying the chaotic ionization of an electron wave packet.  The experimental protocol is based on the theoretical observation \cite{Burke09} (see also Ref.~\cite{Mitchell04a}) that the ionization mechanism can be explained in terms of a homoclinic tangle and its corresponding turnstile~\cite{MacKay84,Wiggins92}.  The turnstile is a structure within phase space that promotes the electron from a bound to an unbound state, and thus serves as the critical step in the ionization process.  Turnstiles have been theoretically applied to chaotic transport in a wide variety of physical systems~\cite{Wiggins92}.  Experimental studies, however, are significantly more limited, with notable examples to chaotic or turbulent fluids~\cite{Solomon96}, including recent work on Lagrangian coherent structures~\cite{Voth02,Shadden06}, and to optical microcavities~\cite{Shim08}.  However, to our knowledge, their structure has not previously been experimentally measured in atomic ionization.  We present experimental data on the observation of the phase-space turnstile in a system consisting of highly excited quasi-one-dimensional Rydberg atoms~\cite{Dunning09} exposed to alternating electric-field pulses, or ``kicks".  The ionization probability depends not only on the electron energy, but, crucially, on the phase-space position of the electronic state with respect to the turnstile at the moment the ionization kicks are applied.  We demonstrate that our measurements are a sensitive probe of the phase-space position and shape of the turnstile.

\begin{figure}
\includegraphics[width=0.49\textwidth]{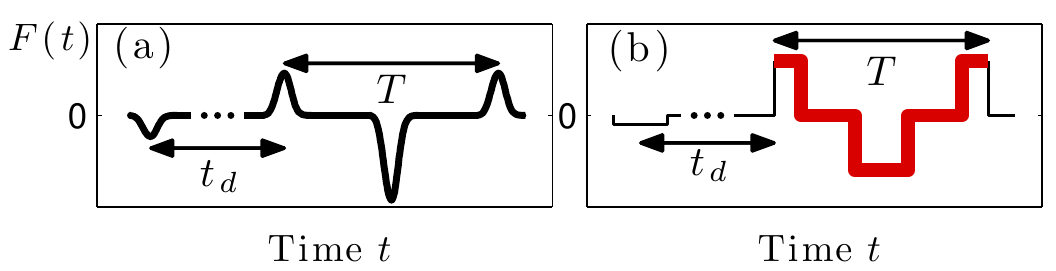}
\caption{\label{fig:kicking_sequence} (Color online.)  (a) The experimental timing sequence for the applied electric force $F(t)$.  The first pulse is the focusing kick.  The remaining three pulses are the ionization kicks. (b) The pulse sequence used in the 1D simulations is a square-wave version of the experimental sequence.  }
\end{figure}

\begin{figure}

  \includegraphics[width=0.5\textwidth]{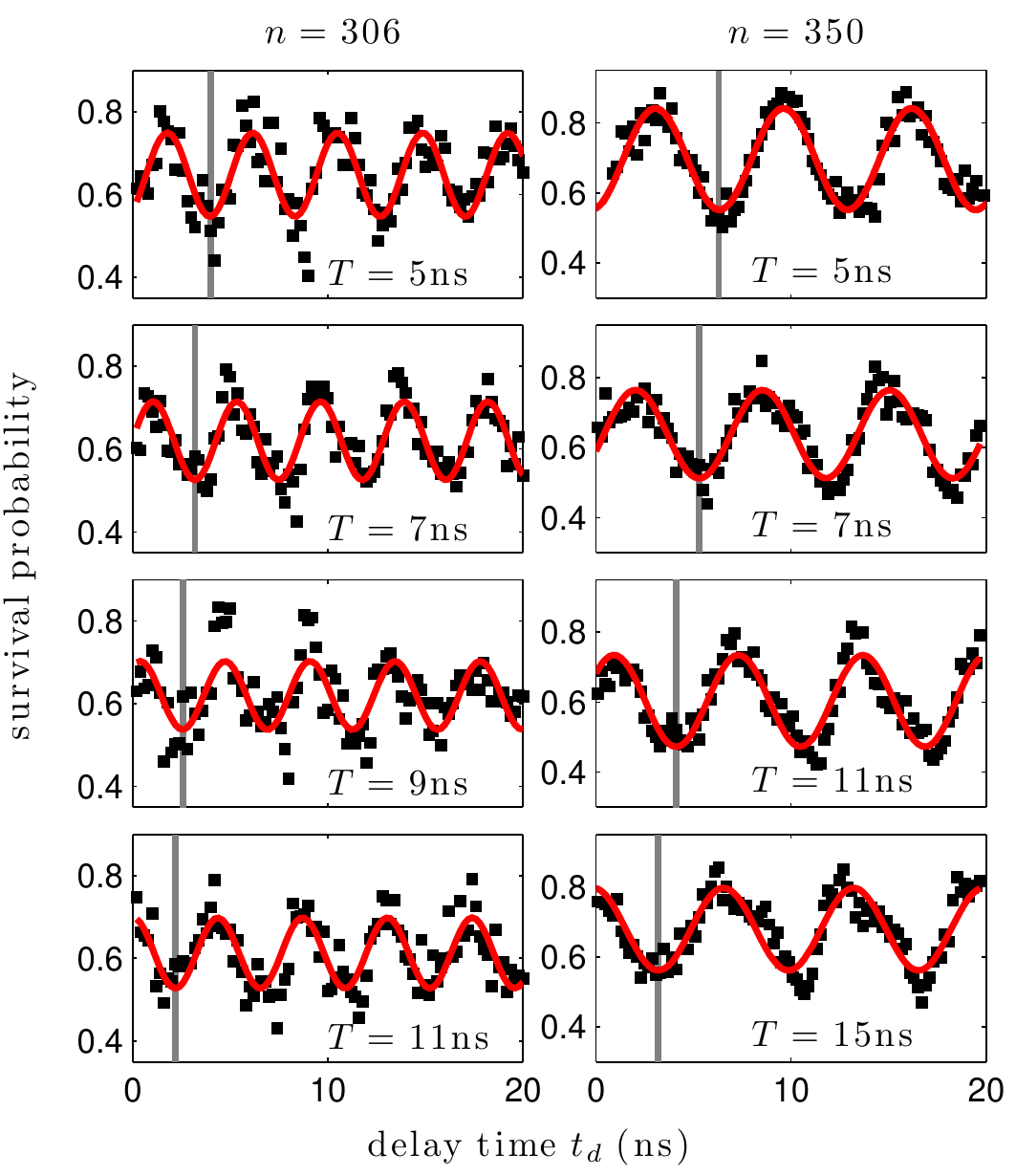} %
  \caption{\label{fig:comparison} (Color online.)  Measured survival probabilities as a function of delay time $t_d$ for the values of peak-to-peak duration $T$ indicated. Squares denote experimental data; red lines are sinusoidal fits \cite{footnote3}.}
\end{figure}

{\bf Experimental protocol and data: } Potassium atoms are first excited to a high principal quantum number ($n \approx 306$ or $n \approx 350$) quasi-one-dimensional Rydberg state using the protocol detailed in Ref.~\cite{Dunning09}.  A small electric field ``focusing'' pulse is applied to the atom that delivers an impulsive kick to the electron of scaled strength $\Delta \tilde{p} = n \Delta p = 0.08$ directed toward the nucleus (Fig.~\ref{fig:kicking_sequence}a). (Tildes denote scaled atomic units: $\tilde{r} = r/n^2$, $\tilde{t} = t/n^3$, $\tilde{E} = n^2E$.) This creates a nonstationary electronic wave packet that undergoes strong periodic focusing near the outer classical turning point.  After some time delay $t_d$, a sequence of three alternating positive and negative ionization kicks is applied (Fig.~\ref{fig:kicking_sequence}a).  These are much stronger than the focusing kick, with $\Delta \tilde{p}= 0.25, -0.5, 0.25$.  The fraction of Rydberg atoms that survive as a function of delay time $t_d$ is measured using field ionization. Figure~\ref{fig:comparison} shows results for both $n \approx 306$ (left column) and $n \approx 350$ (right column).  The delay time $t_d$ is increased from 0\,ns to 20\,ns, and the peak-to-peak duration $T$ of the ionization sequence is adjusted between 5\,ns and 15\,ns.  The kick durations are all fixed at 600\,ps.

\begin{figure}
  \includegraphics[width=0.48\textwidth]{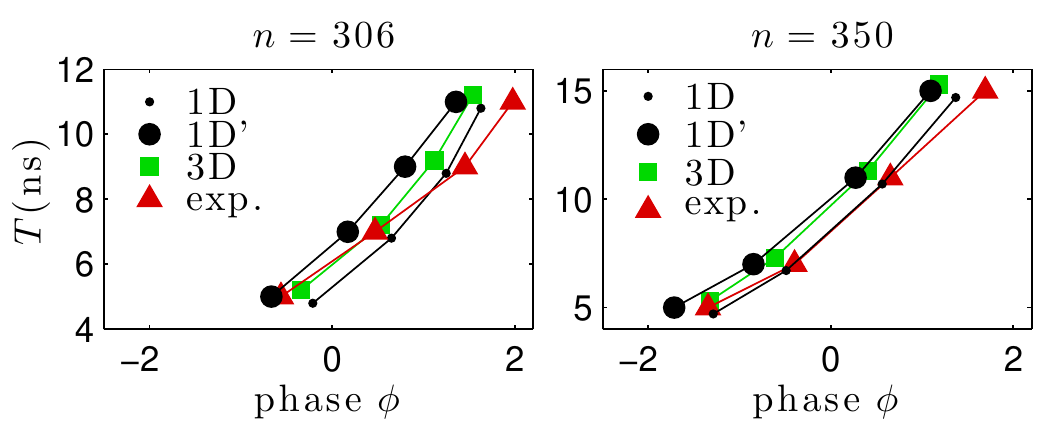} \caption{\label{fig:phase_shift} (Color online.)  Phase shift as a function of $T$.  Small shifts in $T$ are applied to 1D and 3D data to visually separate data markers \cite{footnote3}.}  \end{figure}

The data in Fig.~\ref{fig:comparison} show a clear periodic behavior.  This is elucidated by fitting to a sinusoid $P(t_d)=P_0+A \sin(2 \pi t_d/T_0+\phi)$ and extracting the fitting parameters $P_0$, $A$, $T_0$, and $\phi$.  The first three parameters are recorded in Table \ref{tab:coefficients}; the last is plotted in Fig.~\ref{fig:phase_shift}.  The following features stand out.  (i) The oscillation period $T_0$ is nearly independent of $T$ and matches the classical Kepler period $T_K$ of the original state; for $n=306$, $T_K = 4.35$ns; for $n=350$, $T_K = 6.52$ns.  (ii) The parameter $P_0$ is the survival probability averaged over $t_d$.  It and the oscillation amplitude, $A$, do not depend strongly on $T$.  (iii) The phase shift $\phi$ varies considerably with $T$, as seen in the shift of the vertical lines in Fig.~\ref{fig:comparison}.

\begin{table}
\centering
      \begin{tabular} {|c |c |c |c |c|c|c|c|c|c|c|c|} \hline             
                   $n$ & $T$ &\multicolumn{4}{c|}{$P_0$}&\multicolumn{3}{c|}{$A$}&\multicolumn{3}{c|}{$T_0$(ns)} \\
 \hline \hline  
  \cline{3-12}     &     & exp            &1D             &3D            & 1D'                &exp            &1D                &3D                  &exp                &1D               &3D  \\      
  \cline{3-12}  
306            & 5          &0.648       & 0.647          &0.720       & 0.633            & 0.10        & 0.29          &0.23           & 4.363           & 4.369          & 4.357      \\
                   & 7          &0.620       & 0.636          &0.668       & 0.630           &0.09         & 0.26          &0.22           & 4.286           & 4.419          & 4.348      \\
                   & 9          &0.620       &  0.659         &0.668       & 0.663           &0.08         & 0.22          &0.21           & 4.357           & 4.459           & 4.378       \\
                   &11        &0.613        &  0.688         &0.697       & 0.693            & 0.09        & 0.17          &0.17           & 4.348           & 4.514           & 4.466       \\ \hline
350            & 5          &0.697        & 0.702         &0.760        & 0.692           & 0.14       &0.35           &0.29           & 6.572           & 6.559           & 6.742        \\
                   & 7          &0.638        & 0.648         &0.703        & 0.623           & 0.13       & 0.37          &0.30            & 6.504          & 6.566           & 6.749        \\
                   & 11        &0.603        & 0.633         &0.664       & 0.624            &0.13        & 0.32          &0.27           & 6.372           & 6.607           & 6.684        \\
                   & 15        &0.679        & 0.670         &0.678       &0.669             &0.12        & 0.24          &0.22           & 6.663           & 6.771           & 6.785        \\  \hline

\end{tabular}
\caption{Fit parameters $P_0$, $A$, and $T_0$ from the experiment (exp), 1D and 3D simulations, and a 1D lobe analysis (1D'). } \label{tab:coefficients}
\end{table}

A classical trajectory Monte Carlo (CTMC) simulation is sufficient to reproduce the essential experimental signatures.  We use a one-dimensional hydrogenic model \cite{footnote1} with the square-wave forcing shown in Fig.~\ref{fig:kicking_sequence}b.  The fitting parameters derived from this model are included in Table~\ref{tab:coefficients} and Fig.~\ref{fig:phase_shift} under the label ``1D'' and are in good agreement with the experimental data.  As shown below, these parameters reflect the underlying phase-space geometry.  The amplitude $A$, however, depends more strongly on the details of the electronic initial state, and the experimental oscillation amplitude is somewhat less than the model predictions.  As a consistency check the results of more involved three-dimensional CTMC simulations are also included.

\begin{figure}
\includegraphics[width=0.45\textwidth]{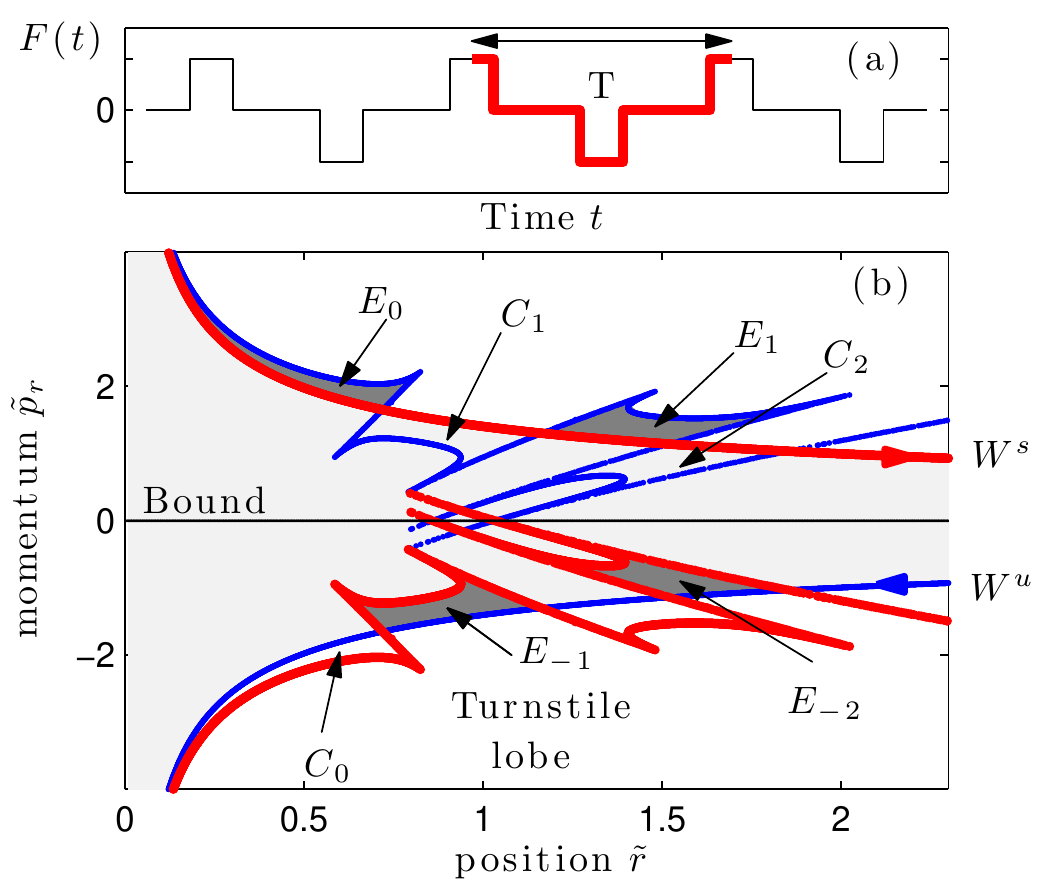} 
\caption{\label{fig:tangle} (Color online.) Panel (a) shows the periodic forcing $F(t)$ used to compute the homoclinic tangle in panel (b).  The stable/unstable manifolds $W^s$/$W^u$ are the thick/thin red/blue lines.  Kicking parameter values are chosen for figure clarity and are not the experimental values used in subsequent figures. }
\end{figure}

{\bf Connection to turnstile geometry:} We now demonstrate how the experimental data reveal the presence of a phase-space turnstile and how the turnstile geometry provides a qualitative and quantitative framework for understanding trends in the data. Consider first the electron dynamics when subjected to periodic forcing, in which the single square-wave forcing cycle of duration $T$ (Fig.~\ref{fig:kicking_sequence}) is replaced by a periodic repetition (Fig.~\ref{fig:tangle}a).  A stroboscopic picture of the dynamics is defined by the Poincar{\'e} map $(r,p_r) \mapsto (r',p_r')$, which takes the radial position and momentum of the electron at a given time and returns their values one forcing period $T$ later.  This map has a fixed point at $r=\infty$ \cite{Burke09} to which stable and unstable manifolds are attached (Fig.~\ref{fig:tangle}b).  (A stable/unstable manifold $W^s$/$W^u$ consists of those phase-space points that converge to the fixed point in the positive/negative time direction~\cite{Wiggins92}.)  The stable and unstable manifolds together form a \emph{homoclinic tangle}~\cite{Wiggins92}(a ``broken separatrix''), which defines the inner gray zone (roughly the ``bound'' electron states) and the outer white zone (the ``ionized'' states.)  The tangle also defines regions called lobes, which fall into two categories, those that govern electron capture ($C_n$) and those that govern electron escape/ionization ($E_n$).  Under the forcing dynamics, $E_{n} \mapsto E_{n+1}$ and $C_{n} \mapsto C_{n+1}$.  The critical step for ionization is the $E_{-1} \mapsto E_0$ transition as this promotes electron states from bound to ionized.  An analogous process $C_{0} \mapsto C_{-1}$ governs capture, and collectively this mechanism is called a phase-space turnstile~\cite{MacKay84}.  Being only interested in ionization here, we call $E_{-1}$ the ``turnstile lobe''.  Its size, shape, and position govern the electron survival probability \cite{footnote2}.

\begin{figure*} 
\includegraphics[width=0.95\textwidth]{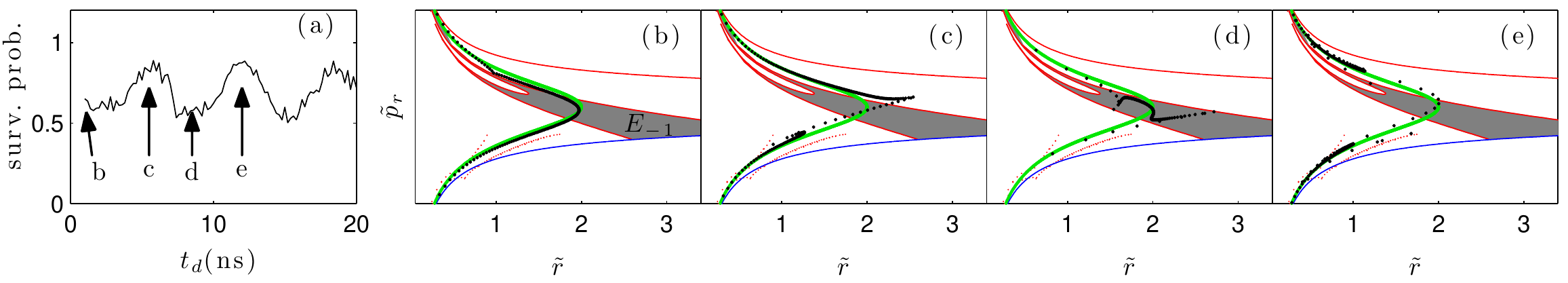}
\caption{\label{fig:focusing} (Color online.) (a) Experimental survival probability for $n=350$, $T=5$ns \cite{footnote3}.  (b)--(d).  Distribution of the classical ensemble (dots) relative to $E_{-1}$, measured at the times indicated.  The thick (green) curve is the $\tilde{E}=-0.5$ energy shell.} \end{figure*}

\begin{figure}
  \includegraphics[width=0.48\textwidth]{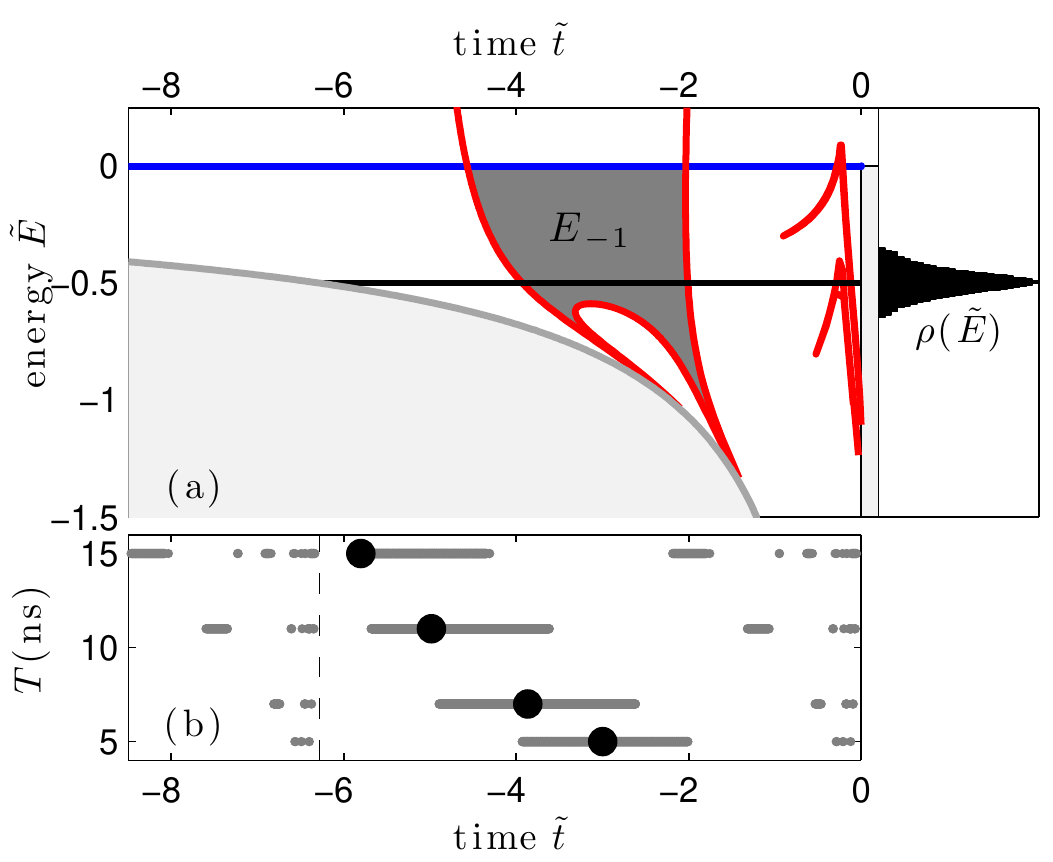} \caption{\label{fig:lobe_and_e} (Color online.)  Panel (a) shows the $E_{-1}$ lobe in energy-time coordinates.  $\rho(\tilde{E})$ illustrates the distribution of energies in the focused ensemble.  Panel (b) shows the intersection segments of the $E_{-1}$ lobe with the $\tilde{E} = -0.5$ line for different $T$ values.  The dashed line is positioned at the negative Kepler period $-2\pi$ and the rightmost segments are repeated on the left taking advantage of the periodic boundary conditions.  The large dots are the average positions of the segments.} \end{figure}

Following the focusing kick, the electronic state is no longer stationary.  The corresponding classical ensemble has a narrow energy distribution $\rho(\tilde{E})$ (Fig.~\ref{fig:lobe_and_e}) centered at $\tilde{E} = -0.5$ with $\Delta \tilde{E} = 0.053$.  It forms a partially localized, or ``focused'', ensemble that hugs and moves along the $\tilde{E} = -0.5$ shell.  (Unless otherwise noted, the remainder uses the $n = 350$ case.)  The observed oscillations in survival probability (Fig.~\ref{fig:comparison}) can now be understood as the ensemble passing in and out of the $E_{-1}$ lobe---survival is higher when the bulk is outside $E_{-1}$ and smaller when inside.  Figure~\ref{fig:focusing} illustrates this with snapshots of the ensemble at a sequence of times: b) Immediately after the focusing kick $t_d = 0$; the ensemble roughly corresponds to the energy shell shifted slightly downward. c) The first survival maximum; the ensemble is almost entirely outside $E_{-1}$, resulting in negligible ionization. d) The first survival minimum; the ensemble has reflected once off the nucleus and now lies almost entirely within $E_{-1}$, resulting in large ionization. e) The second maximum.  Since the ensemble trajectories lie close to the original energy shell, the survival oscillation period $T_0$ is approximately the Kepler period $T_K$.

Next consider the average survival probability $P_0$; $1 - P_0$ can be interpreted as the fraction of time a trajectory spends inside $E_{-1}$, averaged over the focused ensemble.  To visualize this, it is easiest to work in the canonical coordinates $\tilde{E}$ (electron energy $\tilde{p_r}^2/2-1/\tilde{r}$) and $\tilde{t}$ (time to reach the nucleus).  Consider Fig.~\ref{fig:lobe_and_e}a.  The bottom left curve is the negative Kepler period, forming the left phase-space boundary.  It is physically identified with the right vertical boundary line $\tilde{t} = 0$, which represents nuclear impact.  Time evolution in these coordinates consists of uniform motion rightward along horizontal lines; when a trajectory reaches $\tilde{t} = 0$, it jumps back to $\tilde{t} = -\tilde{T}_K$.  The lobe $E_{-1}$ intersects the energy shell $\tilde{E} = -0.5$ in one large segment.  (See Fig.~\ref{fig:lobe_and_e}b, $T = 5$\ ns, for intersection segments.)  The lobe has two ``horns'', which stretch downward, intersecting the left boundary.  From here, they re-emerge as two tendrils at the right boundary (taking advantage of the periodic boundary conditions), which reach up and again intersect $\tilde{E} = -0.5$ in several short segments.  In energy-time coordinates, the fraction of time a trajectory of energy $\tilde{E}$ spends inside $E_{-1}$ equals the total length of the intersection between $E_{-1}$ and the line of constant $\tilde{E}$, divided by $\tilde{T}_K$.  Since $\Delta \tilde{E}$ for the focused ensemble is small, we need only consider $\tilde{E} = -0.5$, with $\tilde{T}_K = 2 \pi$.  Thus, $1-P_0 \approx L/2 \pi$, where $L = 0.308$ is the total length of the intersection segments in Fig.~\ref{fig:lobe_and_e}b, $T = 5$\ ns.  Thereby, $P_0 = 0.692$, as recorded in Table~\ref{tab:coefficients} column 1D', and in excellent agreement with the experimental value $0.697$.

Previous work \cite{Burke09} showed that as $T$ increases (with kick strength and duration fixed), the lobe in Fig.~\ref{fig:lobe_and_e}a shifts leftward.  This is reflected in Fig.~\ref{fig:lobe_and_e}b, which shows the intersections of $E_{-1}$ with the $\tilde{E} = -0.5$ line for increasing values of $T$.  In particular, the primary wide segment shifts left with increasing $T$.  However, another critical effect is the increase in number and length of the shorter segments on the right.  As above, these segments are the intersections with the horns, which have wrapped around.  Physically, the wide segment contains trajectories that strike the nucleus once during the ionization sequence, whereas the shorter segments contain trajectories that strike the nucleus multiple times during ionization.  As more time elapses between ionization pulses, there is more time for trajectories to strike the nucleus, and the relative importance of these ionization pathways increases.  By $T = 15 \text{ns}$, 24\% of the ionizing trajectories have multiple nuclear impacts.  The corresponding $P_0$ values, computed from the interval lengths in Fig.~\ref{fig:lobe_and_e}b, are recorded in Table~\ref{tab:coefficients} column 1D' for varying $T$.  These agree well with the 1D and experimental data.  Were ionization via multiple impacts not included, this agreement would be notably worse, especially for $T = 15 \text{ns}$.

The leftward shift of the intervals in Fig.~\ref{fig:lobe_and_e}b with increasing $T$ explains the phase shift in Fig.~\ref{fig:phase_shift}.  The large dots in Fig.~\ref{fig:lobe_and_e}b are the average positions $\langle \tilde{t} \rangle$ of the escaping points (where the average is taken using the repeated segments on the left.)
To account for the initial position of the focused ensemble, the values of $\langle \tilde{t}\rangle$ are plotted as $-\langle\tilde{t} \rangle -3\pi/2$ in Fig.~\ref{fig:phase_shift}, where they track the changes in phase seen as T increases.  Ionizing trajectories that impact the nucleus multiple times again play an important role, as they shift the average dot leftward in Fig.~\ref{fig:lobe_and_e}b.

{\bf Conclusions:} Present work identifies the turnstile lobe as the critical mechanism for promoting bound electronic states to ionization, the turnstile geometry providing a convenient framework for explaining the experimental results.  More broadly, the results demonstrate that kicked atomic systems provide a convenient laboratory to explore the turnstile mechanism, common to a wide variety of physical systems exhibiting chaotic transport.  

We thank Jeff Mestayer for initial assistance; K.B.~ acknowledges support of the UCOP Presidential Dissertation Year Fellowship; K.M.~acknowledges the support of NSF grant PHY-0748828;  and F.B.D.~acknowledges the support from the Robert A.~Welch Foundation under grant C-0734 and the NSF under grant PHY 0964819.

\bibliography{PRL_kicked_atoms.bib}

\begin{thebibliography}{15}
\bibitem[{foo({\natexlab{a}})}]{footnote3}
\bibinfo{note}{Due to a triggering problem, the phase shown for the $n=350$, $T
  = 5$ns data was derived from a different run, using 67\% smaller kick
  strengths; this is justified by the relatively weak dependence of the phase
  on the kick strength.}
\bibitem[{foo({\natexlab{b}})}]{footnote1}
\bibinfo{note}{Our model ignores core electrons, due to the negligible core
  scattering of the quasi-1D electronic state \cite{Dunning09}.}
\bibitem[{foo({\natexlab{c}})}]{footnote2}
\bibinfo{note}{A slight difference exists between the turnstile defined for
  periodic forcing (Fig.~\ref{fig:tangle}) and that defined for a single
  forcing cycle as in the experiment (Fig.~\ref{fig:lobe_and_e}).}
\end{thebibliography}

\end{document}